\documentclass[a4paper,fleqn,usenatbib]{mnras}
\usepackage[T1]{fontenc}
\usepackage{ae,aecompl}
\usepackage{graphicx}	% Including figure files
\usepackage{subfig}
\usepackage{tabularx}
\usepackage{amsmath}	% Advanced maths commands
\usepackage{amssymb}	% Extra maths symbols
\usepackage{lineno}

\title[Chromospheric differential rotation]{The differential rotation of the chromosphere and the quiet chromosphere in the falling and rising period of a solar cycle}
\author[K. J. Li]{K. J. Li$^{1,2}$ \thanks{E-mail: lkj@ynao.ac.cn} J. C. Xu$^{1}$\\
$^{1}$Yunnan Observatories, Chinese Academy of Sciences, Kunming 650011,  China\\
$^{1}$State Key Laboratory of Space Weather, Chinese Academy of Sciences, Beijing 100190, China
}
\begin{document}
\label{firstpage}
\pagerange{\pageref{firstpage}--\pageref{lastpage}}
\maketitle

\begin{abstract}
The full-disk chromosphere was routinely monitored in the He I 10830\AA\, line at the National Solar Observatory/Kitt Peak from 2004 Nov.  to 2013 March, and thereby, synoptic maps of He I line intensity from Carrington rotations 2032 to 2135 were acquired. They are utilized to investigate the differential rotation of the chromosphere and the quiet chromosphere during the one falling (descending part of solar cycle 23) and the one rising (ascending part of solar cycle 24) period of a solar cycle.
Both the quiet chromosphere and the chromosphere are found to rotate slower and have a more prominent differential rotation, in the rising period of solar cycle 24 than in the falling period of solar cycle 23, and an illustration is offered.
\end{abstract}

\begin{keywords}
Sun: rotation-- Sun: chromosphere-- Sun: activity
\end{keywords}

\section{Introduction}
The solar atmosphere is magnetic, and the magnetic field is the main source of solar activities and changes (Xiang et al. 2014, 2020; Li et al. 2022).
The differential rotation of the solar atmosphere has always been an important field in solar physics and  has been widely investigated due to its intimate relationship to the solar magnetic field through the solar dynamo (Beck 2000; Thompson et al. 2003).
The entire solar atmosphere, from the bottom photosphere layer to the top corona layer,  has been  known to differentially rotate, that is, its  rotation rate is the largest at the equator and decreases with  increasing of latitudes, perhaps due to the effect of the Coriolis force exerted by the rotational motion on the convective motion (Howard 1984;  Chandra et al. 2009, 2010; W\"{o}hl et al. 2010; Vats $\&$ Chandra 2011a, 2011b;  Gigolashvili et al. 2013; Hiremath $\&$ Hegde 2013;  Javaraiah 2013, 2020; Shi $\&$ Xie 2013; Sudar et al. 2014, 2015; Javaraiah $\&$ Bertello 2016; Obridko $\&$ Shelting 2016; Badalyan $\&$ Obridko 2017; Bhatt et al. 2017; Ru\v{z}djak et al. 2017; Xie et al. 2017; Deng et al. 2020;  Singh et al. 2021a, 2021b; Mierla et al. 2020; Edwards et al. 2022;  Poljan\v{c}i\'{c} Beljan et al. 2022). The helioseismic measurements of the inner zonal and meridional flows  are the foundation for understanding  the differential rotation of the solar atmosphere and its relationship with the magnetic fields and the solar dynamo (Howe 2009; Komm et al. 2009).

In the photosphere, the rotation rate of plasma determined by spectroscopic measurements is consistently lower than that of sunspots determined by tracer measurements, and  small sunspots rotate slightly faster than big ones in general (Snodgrass et al. 1984; Jha et al. 2021; Kutsenko 2021). One reasonable speculation for these results is that sunspots of different sizes are rooted at different depths (Beck 2000; Wan et al. 2023). Large-size sunspots are found to repress the differential of rotation in the photosphere
(Braj\v{s}a et al. 1996, 2006; Jurdana-\v{S}epi\'{c} et al. 2011; Li et al. 2020; Wan $\&$ Gao 2022).
One plausible explanation for this result is that appearance of sunspots at low latitudes ($10^{\circ}\sim 35^{\circ}$) should upraise
the rotation profile of photosphere plasma at the latitudes, due to that sunspots rotate faster than the photosphere plasma (Li et al. 2020).

In the chromosphere, short-lived features observed in the Ca II K line rotate at the same velocity as the chromospheric plasma (Antonucci et al. 1979a, 1979b; Xu et al. 2019, 2020), which actually means that small-size magnetic elements rotate at the same rate as the quiet (background) chromosphere, in agreement with Li et al. (2020, 2022). The heating and formation of the quiet chromosphere are believed to be caused mainly by the small-scale magnetic elements whose magnetic flux is in range of $(2.9 - 32.0)\times 10^{18}$ Mx, because the long-term evolution of the quiet chromosphere (intensity values at its full surface) is found to be in anti-phase with the solar cycle (Li $\&$ Feng 2022; Li et al. 2022), as the long-term evolution of the small-scale magnetic elements does (Jin et al. 2011; Jin $\&$ Wang 2012). We note that in this context, ``anti-phase" refers to two variables behaving in an opposite manner such that one reaches its peak, the other is at its minimum, and vice versa. Therefore, the quiet chromosphere rotates at the same rate  as the small-scale magnetic elements do (Li et al. 2020, 2023).  It is speculated that the small scale magnetic elements come from the leptocline (Lefebvre et al. 2009), and it rotates clearly faster than the surface (photosphere) rotation
(Larson $\&$ Schou 2018). Therefore, the quiet chromosphere  and the chromosphere rotate clearly faster than the photosphere.
The quiet chromosphere is found to rotate faster than sunspots at relatively high latitudes of the butterfly pattern of sunspots, but slower at lower latitudes (Li et al. 2020, 2023; Wan $\&$ Gao 2022; Wan et al. 2022), and thus the chromosphere rotates a bit faster than the quiet chromosphere at lower latitudes, but slightly slower  at relatively high latitudes of the butterfly pattern. Therefore, the level of the differentiality of the rotation of the chromosphere is higher than that of the quiet chromosphere, that is, the large-scale magnetic fields strengthen the differential in the chromosphere.  The level of the differentiality of rotation is in order from smallest to largest: the quiet chromosphere, the quiet photosphere, the chromosphere, and sunspots (Li et al. 2023).

The corona is more dynamic than the underlying chromosphere and photosphere, and thus  it is difficult to continue for one rotation period or more to track coronal features. Therefore the results  obtained by different methods for coronal rotations of different features at different time intervals  are sometimes mixed,  and in particular  the results for the differential in coronal rotations are even contradictory to some extent.
Nevertheless, some consensus has been reached. Like the chromosphere, the corona rotates faster and less differentially than the  photosphere  (Mancuso $\&$ Giordano 2011; Mancuso et al. 2020; Sharma et al. 2021, 2023).

The relationship between the rotation of the photosphere and solar activity or the solar cycle is an interesting topic and has been widely studied (Balthasar $\&$ W\"{o}hl 1980; Obridko $\&$ Shelting 2001; Jurdana-\v{S}epi\'{c} et al. 2011; Javaraiah 2013; Wan $\&$ Gao 2022; Wan et al. 2022; Poljan\v{c}i\'{c} Beljan et al. 2022). In the photosphere, the torsional oscillation ($TO$) is the most prominent feature of the rotation of the  atmosphere (plasma), and the high-speed zone of the $TO$ pattern basically coincides with the zone of sunspots' appearance (Howard $\&$ LaBonte 1980; Li et al. 2008). Although the zones both individually exhibit the characteristics of the activity cycle, there may be complex links found  between the rotation of the photosphere and solar activity or the solar cycle in practical studies (Braj\v{s}a et al. 2006; Lekshmi et al. 2018), due to the following factors.  First of all, the rotation rate of the photosphere averaged over latitudes at a certain time varies with the range and number of latitudes considered, that is,  there isn't an exact value of  the differential rotation rate averaged over latitudes at a certain moment.  Some parameters, such as the equatorial rotation rate and the differential of rotation (the latitudinal gradient of rotation), have more suitable practicability to describe the differential rotation than the rotation rate does.
Secondly, in general in the photosphere,  large-scale magnetic fields suppress the differential of rotation rates, while  small-scale magnetic fields seemingly reflect the differential of rotation rates (Li et al. 2013a, 2013b), and thus sunspots of large and small scales may appear at different times and latitudes within a solar  cycle, exerting different effects on the rotation of the photosphere.
In the chromosphere layer, small-scale magnetic fields, which
cause the rotation of the quiet chromosphere to be the same as their own rotation (Li et al. 2022, 2023), are  in anti-phase with the solar cycle (Jin et al. 2011; Jin $\&$ Wang 2012), but the differential of rotation in the quiet chromosphere is recently found to have no relation with the solar cycle (Wan et al. 2023).  The differential of rotation in the chromosphere is however found to have a significantly negative relation with the solar cycle, due to the enhancement  effect of large-scale magnetic fields on the differential of rotation  (Wan $\&$ Gao 2022). During the maximum epoch of a solar cycle, there are more large-scale magnetic fields and stronger enhancement effects, and thus the differential of rotation
is smaller (its absolute value is larger). In the corona layer, the rotation of the atmosphere (structures) is found to be multifariously  related with solar activity or the solar cycle (Vats et al. 1998; Jurdana-\v{S}epi\'{c} et al. 2011; Sharma et al. 2023). Before understanding the relationship of coronal rotations and the magnetic fields, we need to investigate the relation between the formation of the corona and the magnetic fields (Li et al. 2022).

The chromosphere was routinely monitored in the He I 10830\AA\, absorption line at the National Solar Observatory/Kitt Peak from 2004 Nov.  to 2013 March, and synoptic maps in the time interval are achieved.  Li et al. (2020) once investigated the differential rotation of the chromosphere by means of the data.  In this study, the entire data of more than 8 years will be divided into two parts, the descending part of the solar cycle 23 and the ascending part of the solar cycle 24, and then the differential rotation in different phases of a solar cycle will be investigated.

\section{Data and methods}
Data used in this study are synoptic maps of He I line intensity from Carrington rotations 2032 to 2135, which came from daily observations of the full-disk chromosphere in the He I 10830\AA\, line at the National Solar Observatory/Kitt Peak (Livingston et al. 1976; Harvey $\&$ Livingston 1994), in the time interval of 2004 Nov. (later than the maximum time of cycle 23, 2000 July) to 2013 March  (earlier than the maximum epoch of cycle 24, 2014 February).  In all, the data covers 104 synoptic maps. Each synoptic map was gauged across 360 equidistant longitudes which span from $1^{\circ}$  to $360^{\circ}$  and 180 latitudes. The latitudinal measurements took 180 uniform steps based on the sine of the latitude, extending from $-$1 (the south pole) to $+$1 (the north pole). By compiling all the 104 synoptic maps chronologically, we achieved one seamless synoptic image. The intensity of the He I line is in arbitrary but constant unit. Fig. 1 shows the data just within latitudes not higher than $30^{\circ}$. Details of the data may be found in Li et al. (2020). In order to determine periodicity of a time series, the autocorrelation analysis method will be utilized in the following analyses (for details, please refer to Xu $\&$ Gao (2016)), and this method was once used to determine the rotation period of the entire data length by Li et al. (2020), but here the data are divided into two parts.  The minimum epoch of cycle 24 is Dec. 2008 (Carrington rotation 2077), then accordingly, the first part includes Carrington rotations 2032 to 2076, belonging to the descending phase of cycle 23,  and the second is the synoptic images of Carrington rotations 2077 to 2135, belonging to the ascending phase of cycle 24.
Here we take the same processing procedure as Li et al. (2020) did, but the entire data used by Li et al. (2020) are replaced respectively by  the two parts. Resultantly as examples, Fig. 2 shows correlation coefficient of a time series $vs$ itself, varying with their relative phase shifts, in each of the total 4 panels, and two panels in the figure show the calculated results of time series in the ascending phase, while the other two,  those of time series in the descending phase. In the figure, CR means the synodic Carrington rotation period, and 1CR = 27.2753 days.
When the synodic rotation
period (SRP) of a series at a measurement latitude is determined, then the rotation velocity at the latitude can be known:
$SRP\times 360/27.2753\approx 13.1988 SRP$ (in degrees/day; Li et al (2020)). Finally, synodic rotation velocities at all measurement latitudes which are not higher than $30^{\circ}$ are determined, which are shown in Fig. 3, and  meanwhile the correlation coefficient corresponding to the SRP at a measurement latitude is displayed also in the figure, which is statistically significant at the $99\%$ confidence level.

\begin{figure}
\begin{center}
\includegraphics[width=1.01 \textwidth]{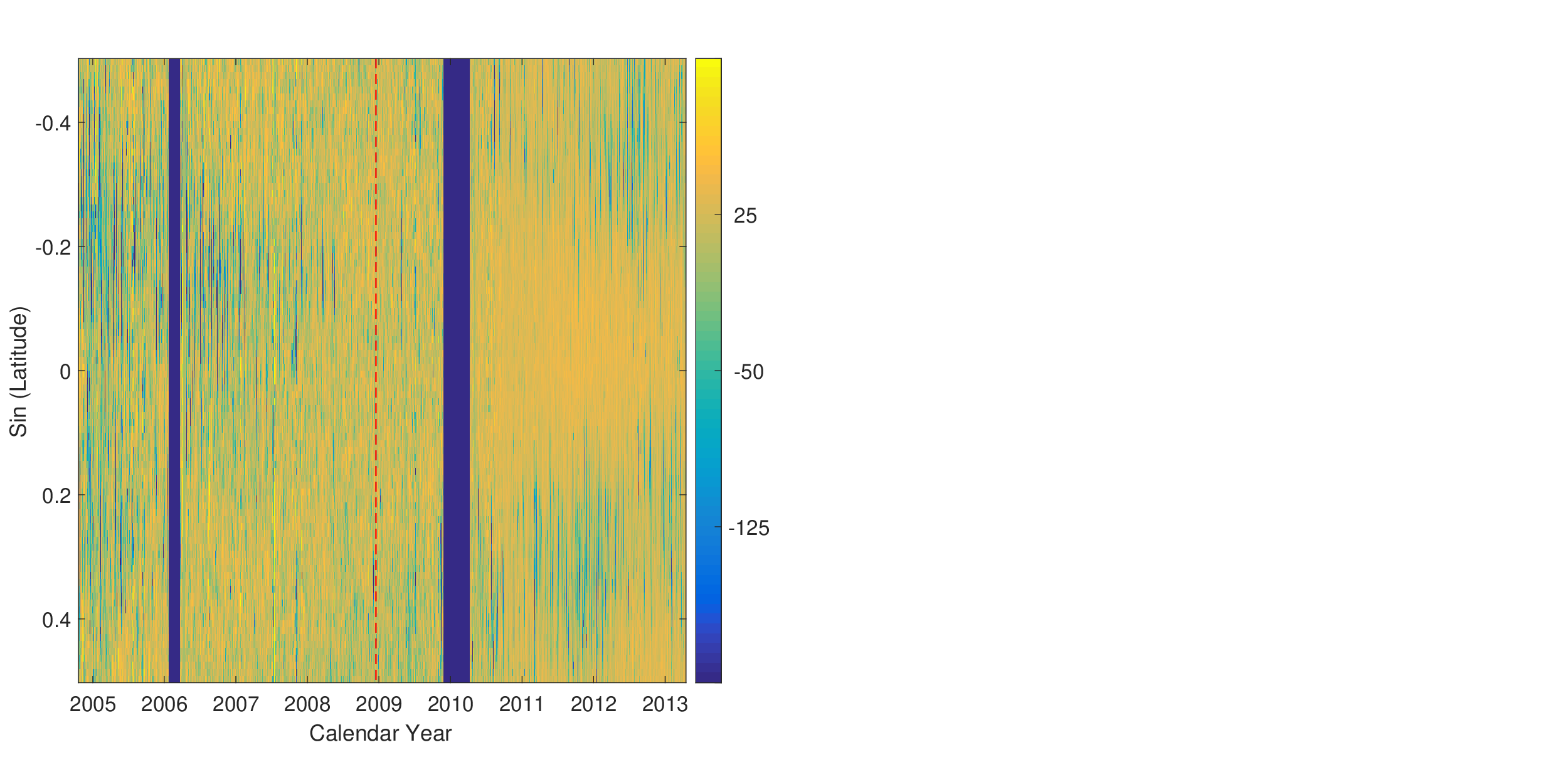}
\caption{Synoptic images (time-latitude diagram)  of He I intensity in the interval of 2004 Nov. to 2013 March, but only latitudes not higher than $30^{\circ}$ are displayed here. The color bar indicates the intensity of the He I line with an arbitrary but constant unit. The two vertical  blue bands  indicates no observational data available, and the red dashed line shows the minimum time of cycle 24.
}\label{}
\end{center}
\end{figure}

\begin{figure}
\begin{center}
\includegraphics[width=1.01 \textwidth]{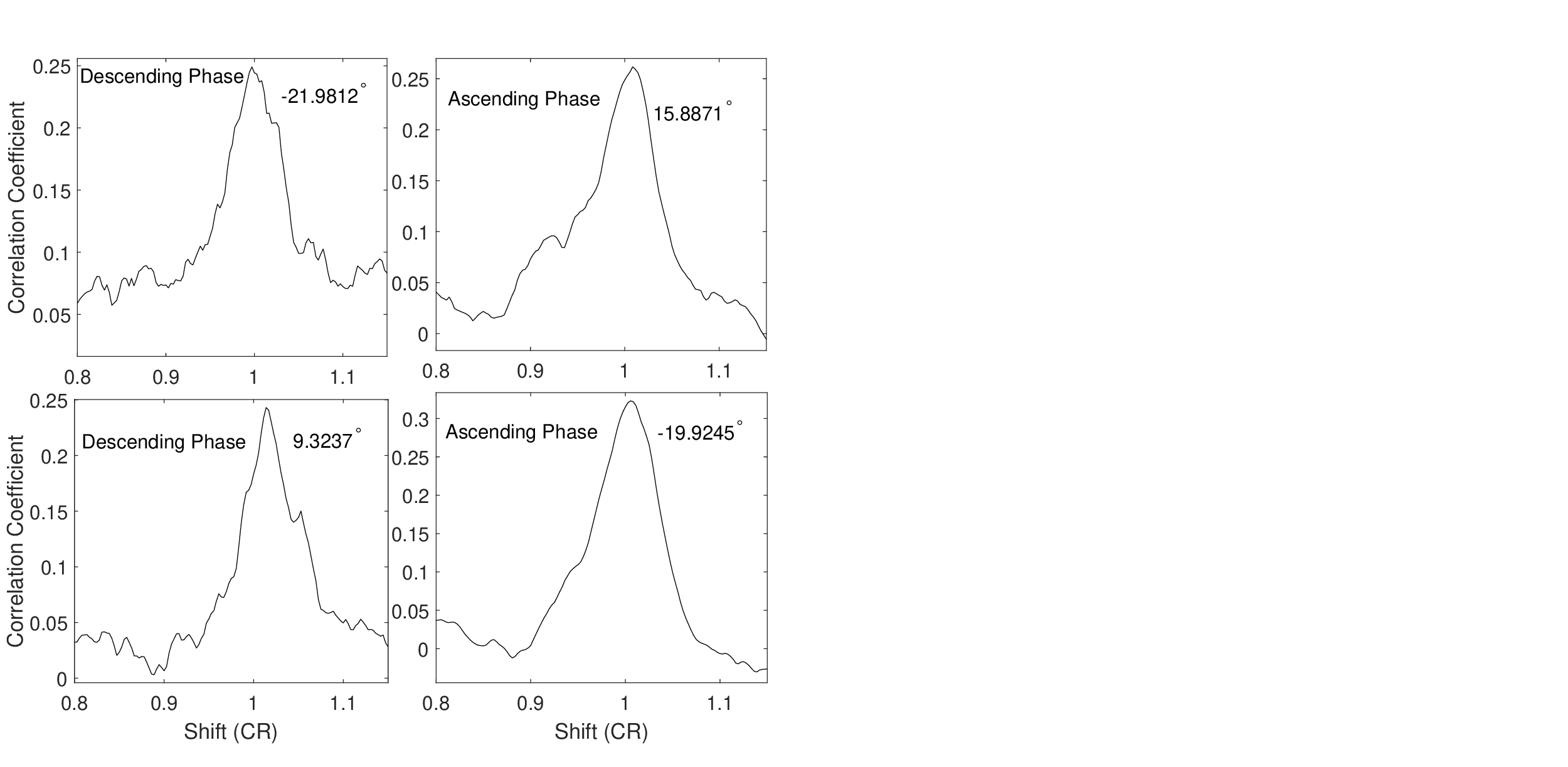}
\caption{Correlation coefficient of a time series of He I line intensity at a certain latitude $vs$ itself, varying with their relative phase shifts. Four panels correspond to four time series, whose latitudes are displayed in the upper right corner of each panel. In a panel, the words ``Ascending Phase"  mean that a time series of He I line intensity  in the ascending phase of solar cycle 23 is used to determine the rotation period of the series, while
the words  ``Descending Phase"  mean that a time series of He I line intensity in  the descending phase of solar cycle 24 is used.
}\label{}
\end{center}
\end{figure}

\begin{figure}
\begin{center}
\includegraphics[width=1.01 \textwidth]{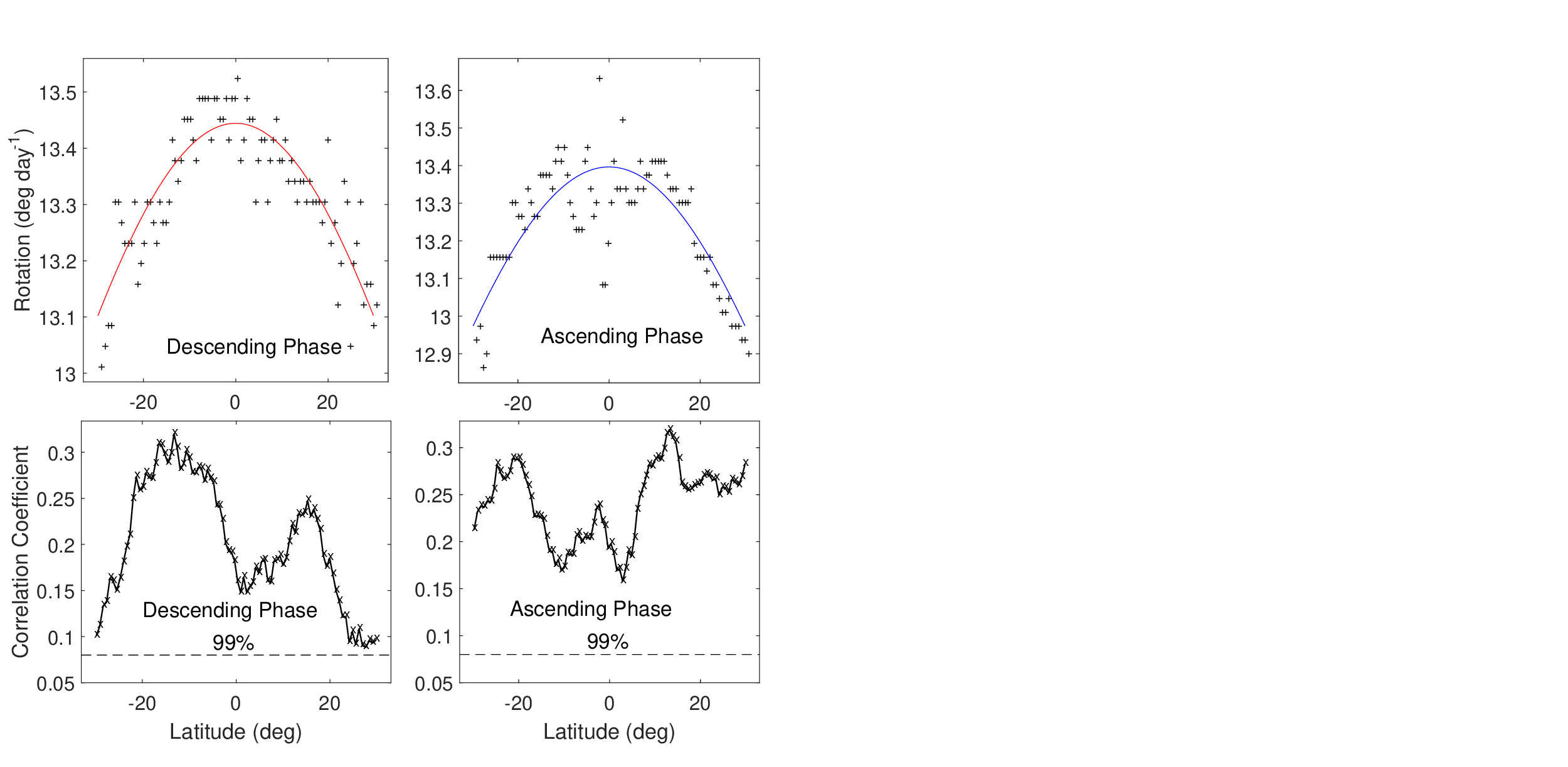}
\caption{Synodic rotation velocities (pluses, top panels) and their corresponding correlation coefficients (crosses in a black solid line, bottom panels) at measurement latitudes which are not higher than $30^{\circ}$, determined respectively in the descending phase (left panels) and the ascending phase (right panels). The red and blue solid lines are the fitting lines of synodic rotation velocities respectively in the descending and ascending phases, and the dashed lines show the $99\%$ confidence level.
}\label{}
\end{center}
\end{figure}

\section{Differential rotation of the  chromosphere}
At low latitudes ($\varphi$) not higher than $30^{\circ}$,  the following theoretical expression is usually utilized to fit observational velocities ($\omega$) of the differential rotation:
$$\omega(\varphi)= A+ B\sin^{2}(\varphi), $$
where the undetermined parameter $A$ is  the rotation velocity at the solar equator, and the undetermined parameter $B$ reflects the differential degree of rotation.  Here the expression is utilized to fit observational rotation velocities, and the bootstrap method (Xu et al. 2019) is used to determine values of the undetermined parameters ($A$ and $B$) and their errors. The differential rotation of the He I chromosphere in the descending phase of cycle 23 is determined to be: $$\omega(\varphi)= 13.4444\pm 0.0091+(-1.3825\pm0.1020) \sin^{2}(\varphi),$$  and that in the ascending phase of cycle 24, $$\omega(\varphi)= 13.3967\pm 0.0190+(-1.7091\pm0.1493) \sin^{2}(\varphi).$$
The parameters $A$ and $B$ in the descending phase are different from those in the ascending phase. Specifically in the descending phase, the rotation at the equator is faster, but differential degree is lower, than  in the ascending phase.

These two fitting lines are put together and shown in Fig. 4.
The difference of rotation velocity at a measurement latitude respectively in the descending and ascending phases is also displayed in the figure. Of the 90 measurement latitudes,  rotation velocity determined in the descending phase is  equal to that in the ascending phase at 13 latitudes, larger at 57 latitudes, and smaller at 20 latitudes.
The Wilcoxon test method for paired observations (Hollander $\&$ Wolfe 1973) is utilized to examine statistical significance of the difference between these two velocity distributions. The calculated Z-value for the $+$ sign is 4.22, which is larger than the tabulated critical value (2.58) at the 0.01 significance level.
Fig. 4 also shows the error ($1\sigma$) lines of these two fitting lines, and the difference between the fitting lines is significant at the $1\sigma$ level (Braj\v{s}a et al. 1995, 1999).
Therefore generally, the rotation velocity in the descending phase is  greater than that in the ascending phase.

\begin{figure}
\begin{center}
\includegraphics[width=1.01 \textwidth]{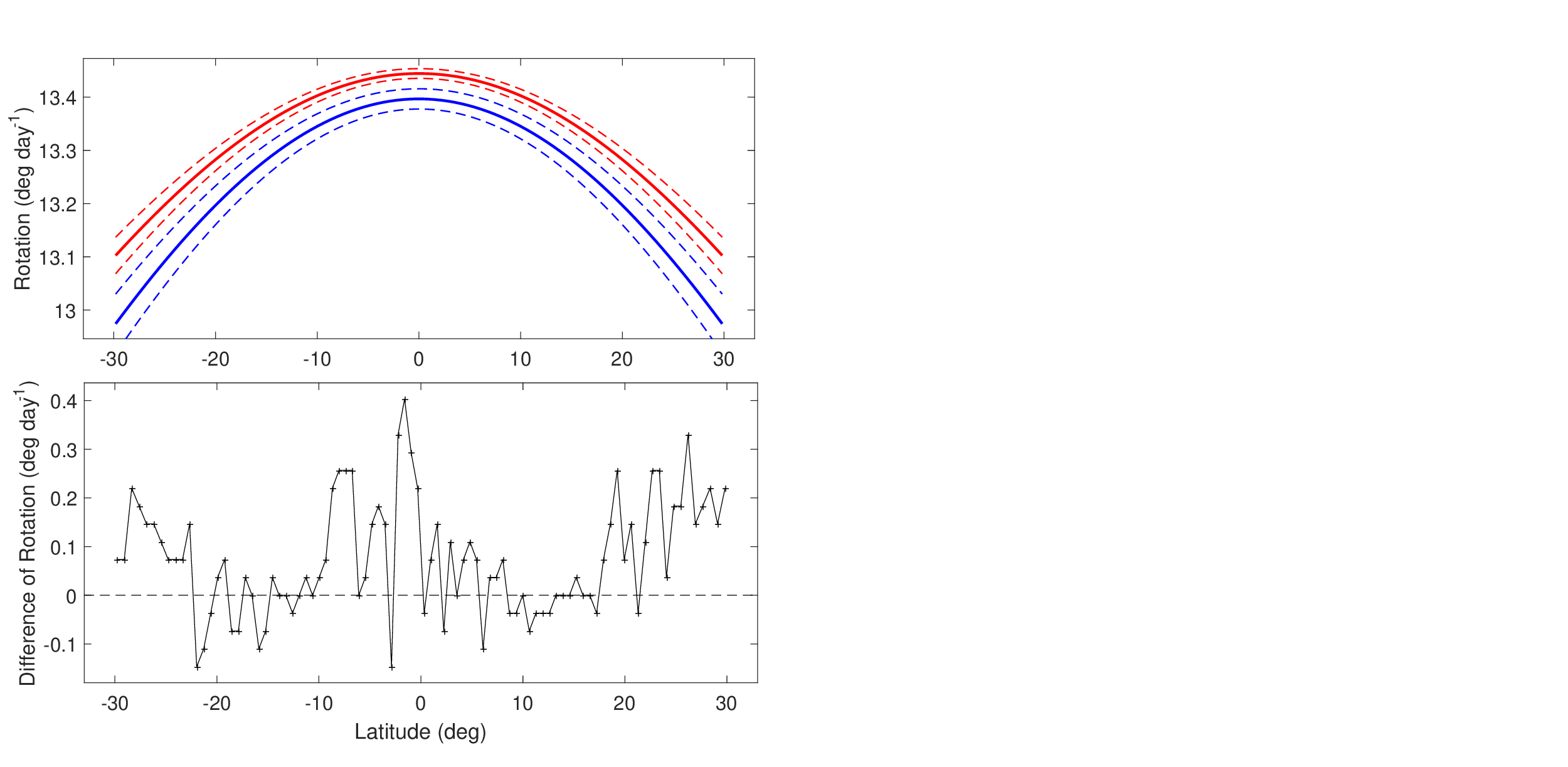}
\caption{Top panel: the fitting line of synodic rotation velocities respectively in the descending phase of cycle 23 (the red solid thick line) and the ascending phase of cycle 24 (the blue solid thick line).
The red (blue) dashed thin line is the error line of the red (blue) solid thick line.
Bottom panel:  difference (crosses) of the value respectively in the red and blue lines at a measurement latitude.
}\label{}
\end{center}
\end{figure}

\section{Differential rotation of the  quiet chromosphere}
Those intensities not less than -10  can be  approximated as coming from the quiet
chromosphere (Li et al 2020). Note that the He I line intensity in synoptic maps is represented in arbitrary but constant unit. Therefore in order to investigate the differential rotation of the quiet chromosphere, it is them that are considered in the following analysis. We repeat the above analysis, but intensities in the chromosphere are replaced by those in the quiet chromosphere.
Here  as four examples, Fig. 5 shows four calculated correlation-coefficient profiles at the same latitudes as Fig. 2, and
Fig. 6 is similar to Fig. 3, showing the determined synodic rotation velocities and the corresponding correlation coefficients. These coefficients are significant at the $95\%$ confidence level. The expression is also utilized to fit observational rotation velocities, and  the differential rotation of the quiet chromosphere in the descending phase is: $$\omega(\varphi)= 13.4011\pm 0.0183+(-0.9121\pm0.2332) \sin^{2}(\varphi),$$ and that in the ascending phase, $$\omega(\varphi)= 13.3342\pm 0.0210+(-1.3492\pm0.1729) \sin^{2}(\varphi).$$
For the quiet chromosphere, the rotation at the equator is faster, but differential degree is lower, in the descending phase than  in the ascending phase.

\begin{figure}
\begin{center}
\includegraphics[width=1.01 \textwidth]{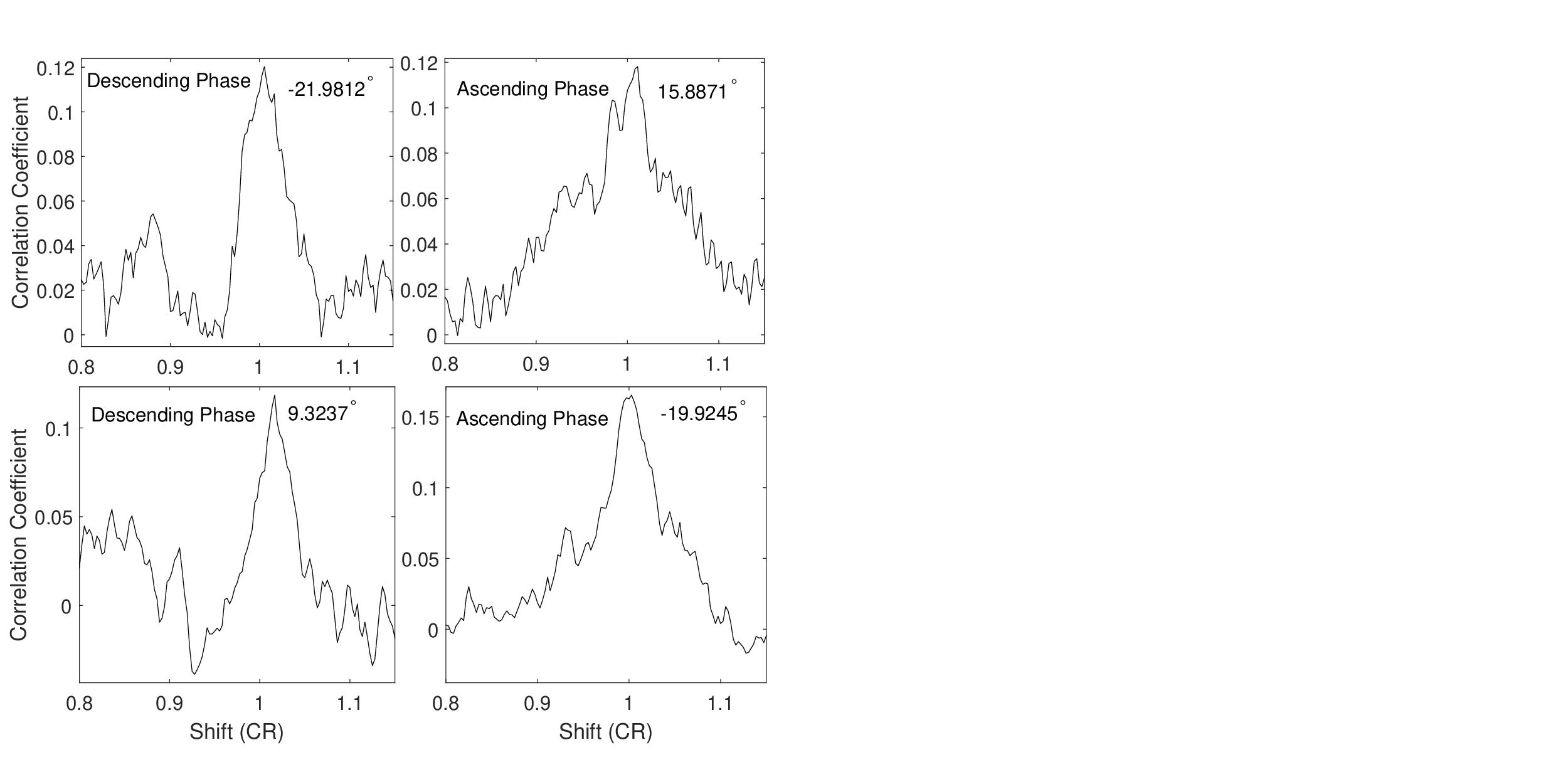}
\caption{The same as Figure 2, but just those intensities which are not less than -10 are considered, that is, the quiet chromosphere is considered instead of the chromosphere.
}\label{}
\end{center}
\end{figure}

\begin{figure}
\begin{center}
\includegraphics[width=1.01 \textwidth]{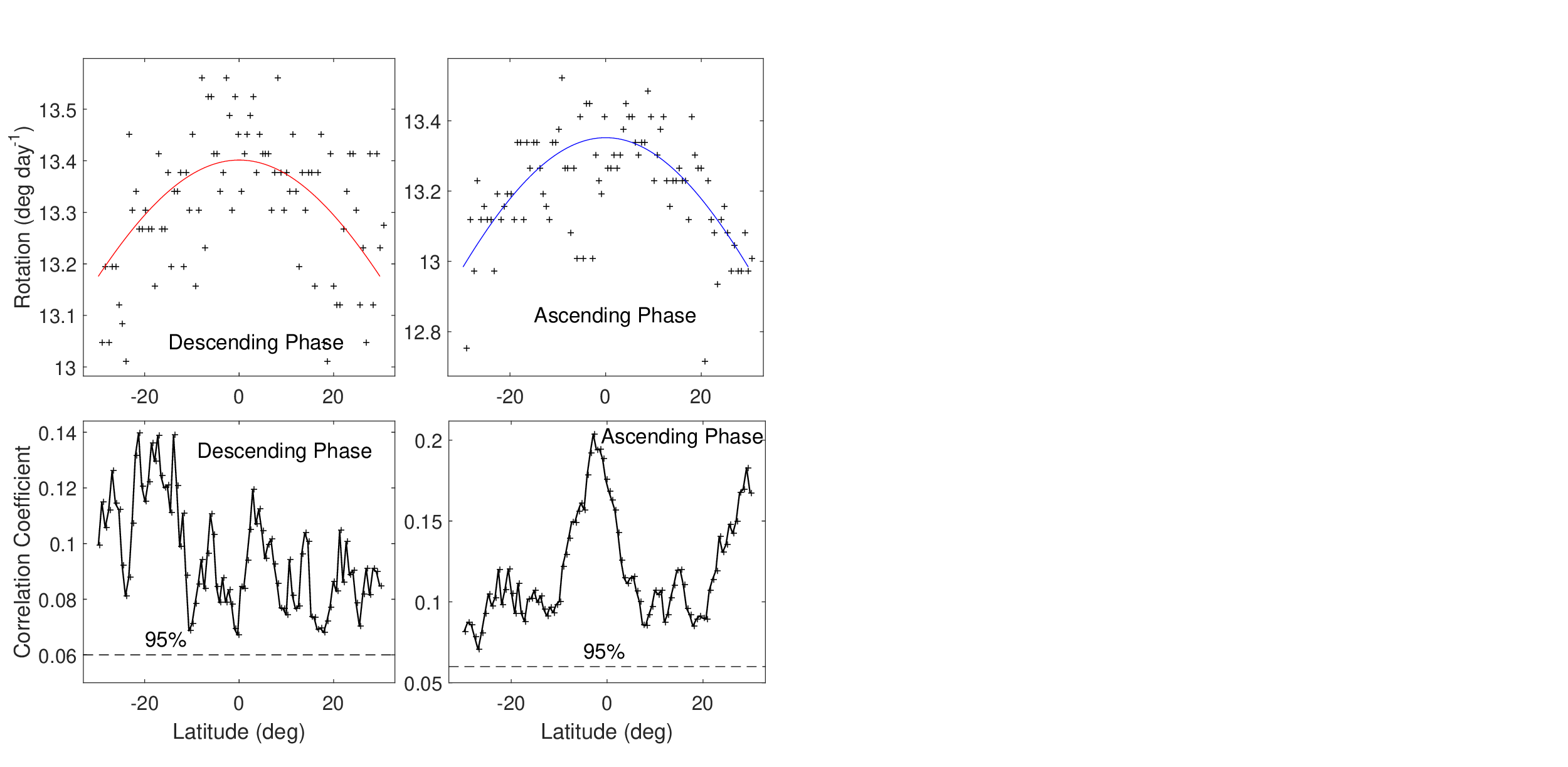}
\caption{The same as Figure 3, but just those intensities which are not less than -10 are considered, that is, the quiet chromosphere is considered instead of the chromosphere.
}\label{}
\end{center}
\end{figure}

These two fitting lines and their error lines are shown together  in Fig. 7, and displayed also in the figure is the difference of rotation velocity respectively in the descending and ascending phases. Rotation velocity in the descending phase is larger at 61 latitudes and smaller at 21 latitudes than that in the ascending phase.
Similarly, the Wilcoxon test method  (Hollander $\&$ Wolfe 1973) is utilized to examine statistical significance of the difference between these two velocity distributions. The calculated Z-value for the $+$ sign is 4.41, which is larger than the critical value
(2.58) at the 0.01 significance level. As the figure displays,
the difference between the fitting lines is significant at the $1\sigma$ level (Braj\v{s}a et al. 1995, 1999).
Therefore in the quiet chromosphere, the rotation velocity in the descending phase is generally greater than that in the ascending phase.

\begin{figure}
\begin{center}
\includegraphics[width=1.01 \textwidth]{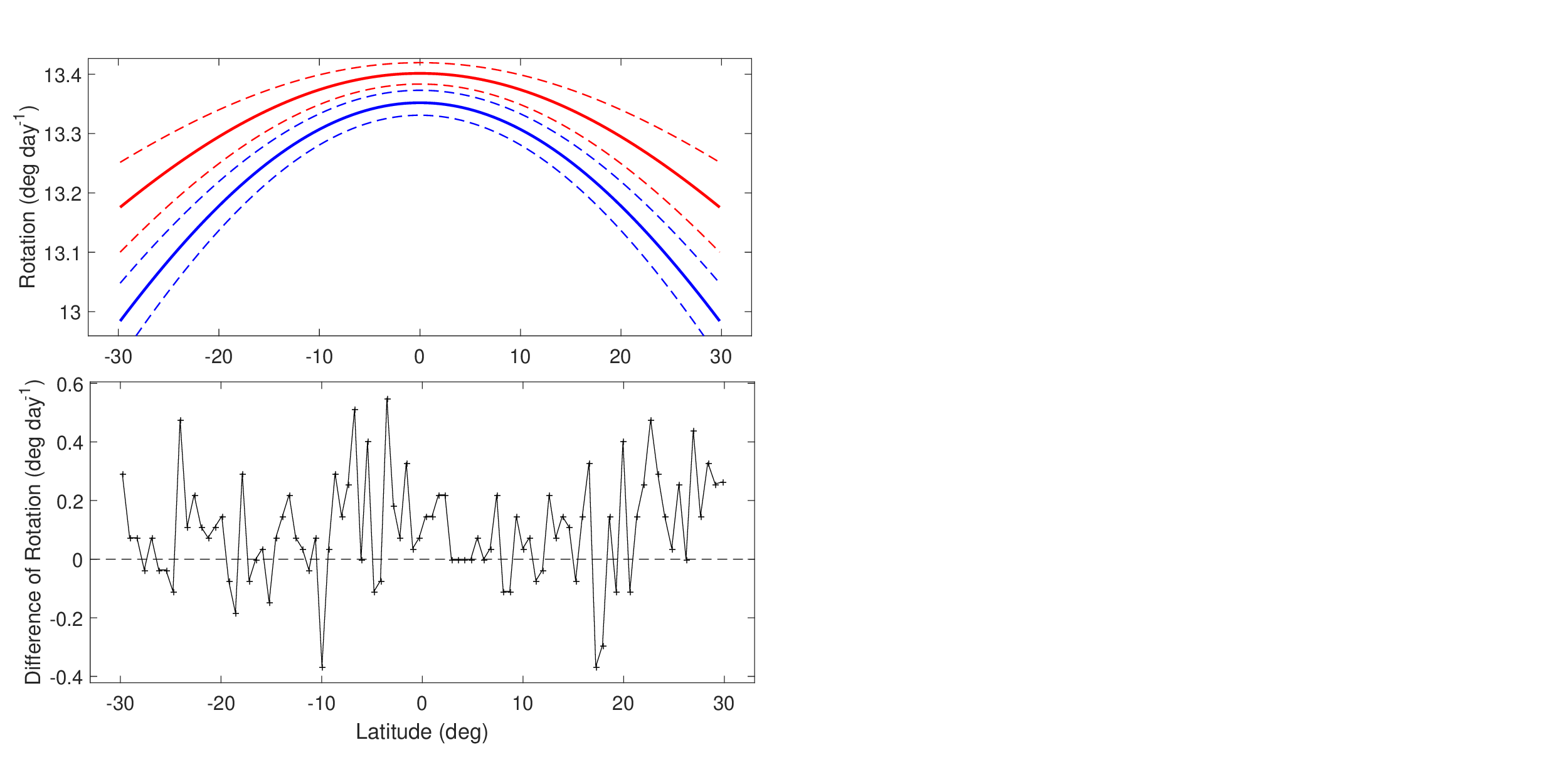}
\caption{The same as Figure 4, but just those intensities which are not less than -10 are considered, that is, the quiet chromosphere is considered instead of the chromosphere.
}\label{}
\end{center}
\end{figure}

\section{Comparison of the differential rotation of the  chromosphere and the quiet chromosphere}
The two fitting lines in the descending phase, corresponding to the differential rotation of the chromosphere and that of the quiet chromosphere,  are displayed together in the top panel of Fig. 8, and the fitting lines in the ascending phase are displayed together in the bottom panel of the figure.
The differential rotation of sunspots  is generally adopted the expression (Howard 1996; Komm et al. 2009):
$$\omega(\varphi)= 13.490 - 2.875\sin^{2}(\varphi)$$ for the synodic rotation, and it is also plotted in the figure.
The commonly used expression for the differential  rotation of the photosphere is (Snodgrass et al. 1984; Komm et al. 2009):
$$\omega(\varphi)= 13.127 - 1.698  \sin^{2}(\varphi)- 2.346  \sin^{4}(\varphi)$$ for the synodic rotation, and  it is drawn in the figure as well.
As the figure indicates, (1) the chromosphere and the quiet chromosphere rotate clearly faster than the photosphere; (2) the chromosphere
rotates faster than the quiet chromosphere at much low latitudes (less than about $20^{\circ}$), but slower at middle latitudes; and (3) sunspots rotate faster than the chromosphere and the quiet chromosphere at much low latitudes (less than about $15^{\circ}$), but slower at latitudes higher than  about $18^{\circ}$.

\begin{figure}
\begin{center}
\includegraphics[width=1.01 \textwidth]{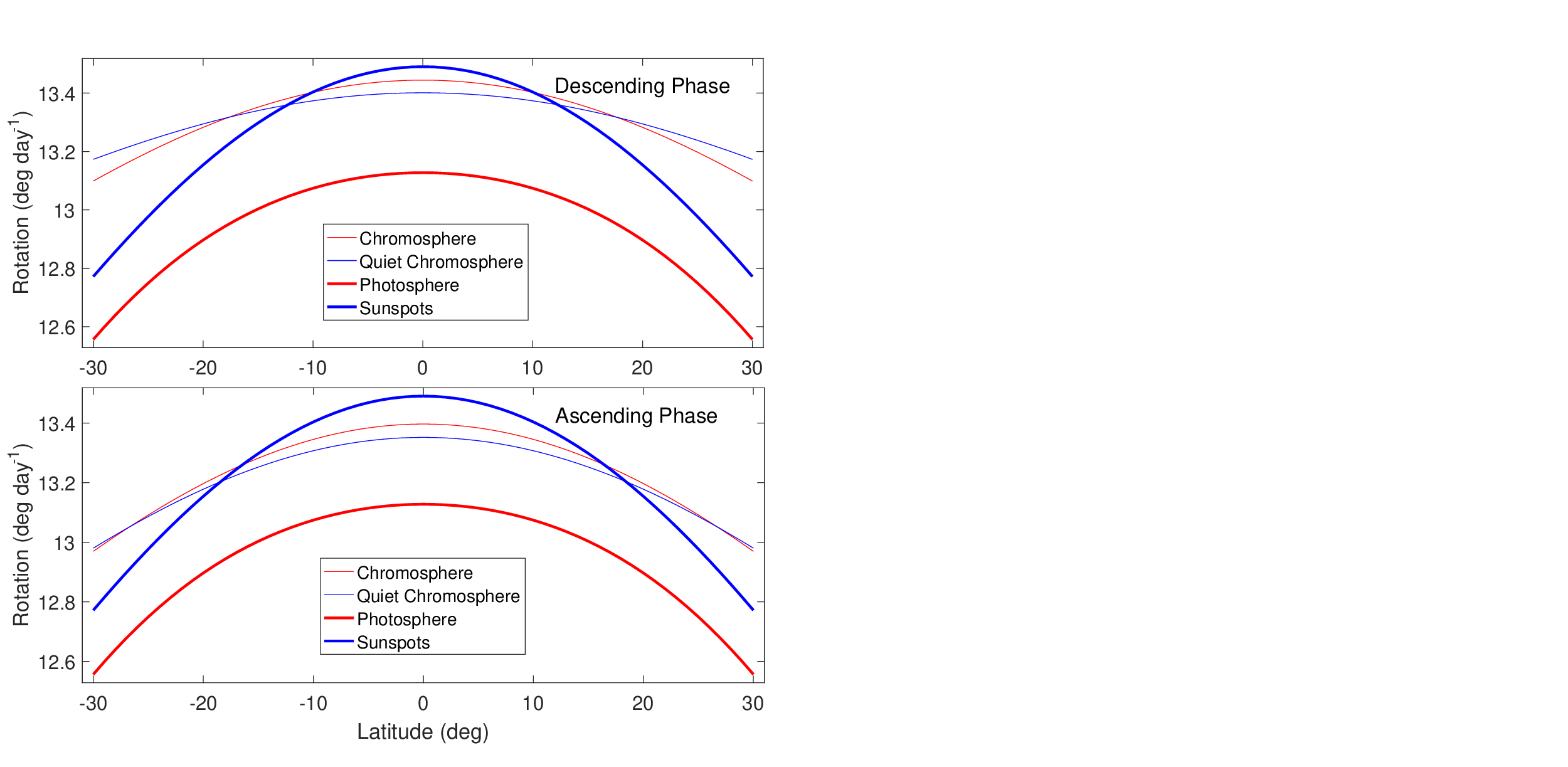}
\caption{Top panel: the fitting line of the differential rotation in the descending phase of cycle 23 respectively for the chromosphere (the thin red line) and the quiet chromosphere (the thin blue line) from the present work. Bottom panel: the fitting line of the differential rotation in the ascending phase of cycle 24 respectively for the chromosphere (the thin red line) and the quiet chromosphere (the thin blue line) from the present work. In both panels, the thick red line displays the differential rotation of the photosphere (adopted from Snodgrass et al. 1984 and Komm et al. 2009), while the differential rotation of sunspots is shown by  the thick blue line (adopted from Howard 1996 and Komm et al. 2009).
}\label{}
\end{center}
\end{figure}

Here a classical  relation between synodic ($\omega_{syn}$) and sidereal $\omega_{sid}$ rotation velocities
is used (Braj\v{s}a et al. 1999; Skoki\'{c} et al. 2014): $\omega_{syn}$=$\omega_{sid}$ - 0.9856 ($deg$  $day^{-1}$), and it is convenient for us to investigate the differential rotation of the chromosphere and the quiet chromosphere during the one falling and one rising period of a solar cycle.
Of course, a precise transformation should  refer to  Skoki\'{c} et al. (2014),
due to the Earth's ellipticity orbit and temporal variation of solar rotation axis to the ecliptic (Graf 1974; Ro\v{s}a et al. 1995; Braj\v{s}a et al. 2002).

\section{Conclusions and discussion}
In this study, the synoptic images of He I 10830\AA\, line intensity  from Carrington rotations 2032 to 2135 are used to explore the differential rotation of both the chromosphere and the quiet chromosphere respectively during the ascending and descending phase of a solar cycle, and some interesting results are obtained.
We note that although we speculate our results should also be applicable to other cycles, the limitations of the available data necessitate more extensive analysis for a definitive generalisation.

Based on this study and Li et al (2020), the following results hold true whether in the rising period, the falling period of a solar cycle,  or a solar cycle. The differential rotation of both the chromosphere and the quiet chromosphere is significantly greater than that of the photosphere; the quiet chromosphere rotates slightly slower than the chromosphere at lower latitudes around the equator, but  a little faster at latitudes higher than $\sim 20^{\circ}$; and sunspots rotate a bit faster than the chromosphere and the quiet chromosphere at latitudes lower than $\sim 15^{\circ}$, but explicitly slower at latitudes higher than $\sim 18^{\circ}$. These results were reasonably explained by Li et al. (2020).  It is mainly the small-scale magnetic fields (SMFs) whose magnetic flux is in the range of (2.9--32.0)$\times 10^{18}$ Mx (Jin et al. 2011; Jin $\&$ Wang 2012) that make the rotation of the quiet chromosphere different from that of the photosphere, and then further, sunspots make the rotation of the quiet chromosphere different from that of the chromosphere; therefore, the quiet chromosphere and SMFs are both in anti-phase with the solar cycle  (Li et al 2022, 2023; Li $\&$ Feng 2022).

The quiet chromosphere is found to rotate faster in the falling period than in the rising period of a solar cycle, but its differential degree of rotation  is smaller in the falling period than in the rising period. Jin and Wang (2012)  gave time - latitude distribution of peak values of the probability distribution functions of latitude for SMFs (the anti-correlated  network elements)  in their Figure 5. The figure indicates  that
SMFs are mainly distributed at latitudes  higher than $\sim 20^{\circ}$ in the falling period of a solar cycle, and at latitudes  lower than $\sim 20^{\circ}$ in the rising period. The rotation velocity of SMFs is greater than that of the photosphere (Xu et al. 2016; Li et al 2020).

It is assumed that the rotation of the quiet chromosphere is caused by the action of SMFs on the rotation of the photosphere (Li et al. 2022), and then the action of SMFs on the rotation of the  photosphere respectively in the falling and rising period is illustrated in Figure 9, based on the time - latitude distribution.
In the falling period, SMFs emerge and act on the rotation of the photosphere from latitudes higher than $\sim 20^{\circ}$, and thus they increase rotation velocities but attenuate the rotation differential of the photosphere.
In the rising period, SMFs emerge and act on the rotation of the photosphere from the latitudes lower than $\sim 20^{\circ}$, and thus they increase rotation velocities and  strengthen the rotation differential of the photosphere. This is why the differential degree of rotation  in the quiet chromosphere is smaller in the falling period than in the rising period.
As for that the quiet chromosphere  rotates faster in the falling period than in the rising period of a solar cycle, perhaps it is because more SMFs appear in the falling period.

\begin{figure}
\begin{center}
\includegraphics[width=1.01 \textwidth]{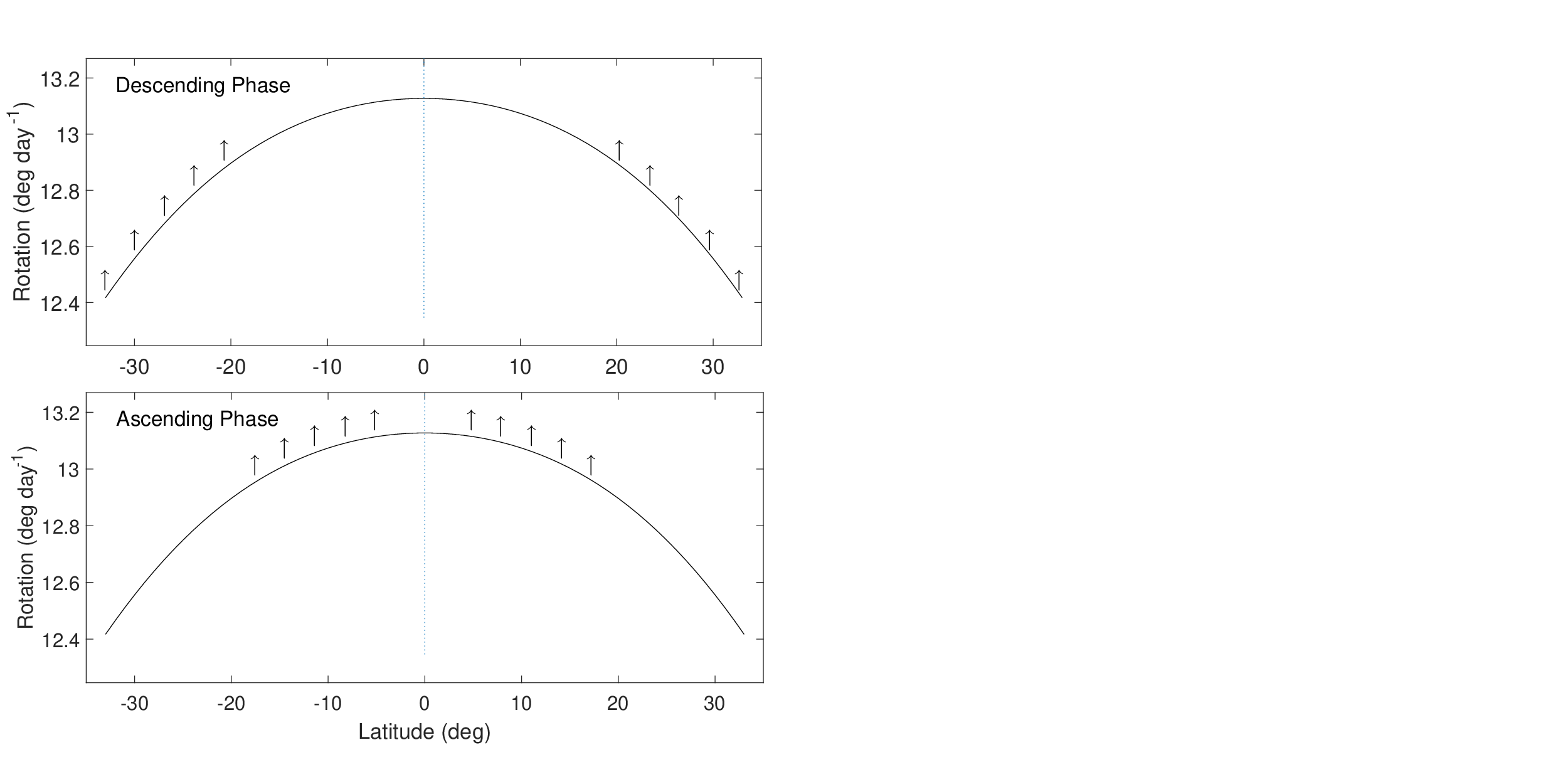}
\caption{Schematic diagram of the effect (up arrows) of small-scale magnetic fields on the differential rotation of the photosphere (the solid line) respectively in the descending (the top panel) and ascending (the bottom panel) phases. Up arrows mean that small-scale magnetic fields increase the rotation velocity of the photosphere.
}\label{}
\end{center}
\end{figure}

Recently, the quiet chromosphere is found to rotate a litter faster than sunspots  at relatively high latitudes of the butterfly diagram of sunspots, but
a bit slower at low latitudes round the equator (Li et al. 2020, 2023). In the falling period of a solar cycle, sunspots are known to generally appear at lower latitudes of the butterfly pattern, and in the rising period sunspots  appear at higher latitudes of the butterfly pattern and low latitudes around the equator, for that some small sunspots of the previous cycle may still appear at low latitudes when a new active cycle begins.
It is assumed that the rotation of the chromosphere is caused by the action of sunspots on the rotation of the quiet photosphere, and then the action of sunspots on the rotation of the quiet chromosphere respectively in the falling and rising period is illustrated in Figure 10.
In the falling period, sunspots emerge  from lower latitudes of the butterfly pattern and act on the rotation of the quiet chromosphere, and thus they increase rotation velocity and  strengthen rotation differential of the quiet chromosphere in general.
In the rising period, sunspots emerge from higher latitudes of the butterfly pattern and low latitudes around the equator and act on the rotation of the quiet chromosphere, and thus they decrease rotation velocities and  strengthen the rotation differential of the quiet chromosphere. This is why rotation velocity in the chromosphere is larger in the falling period than in the rising period.
As for that the rotation differential degree of the chromosphere  is lower   in the falling period than in the rising period of a solar cycle, perhaps it is because the rotation differential degree of the quiet chromosphere  is lower   in the falling period. Of course in the rising and falling period, different degree of sunspot action on the quiet chromosphere may also lead to difference of the chromospheric differential, and the difference is considered minor.

Of course, we need to make it clear that our conclusion is derived using only one rising and falling period of the solar cycle, and thus more rising and falling periods are needed to generalize the conclusion in future.

To sum up, the quiet chromosphere and the chromosphere rotate faster, and the differential degree of their rotation is lower, in the falling period than in the rising period of a solar cycle, due to the effect of the magnetic fields on rotation.

\begin{figure}
\begin{center}
\includegraphics[width=1.01 \textwidth]{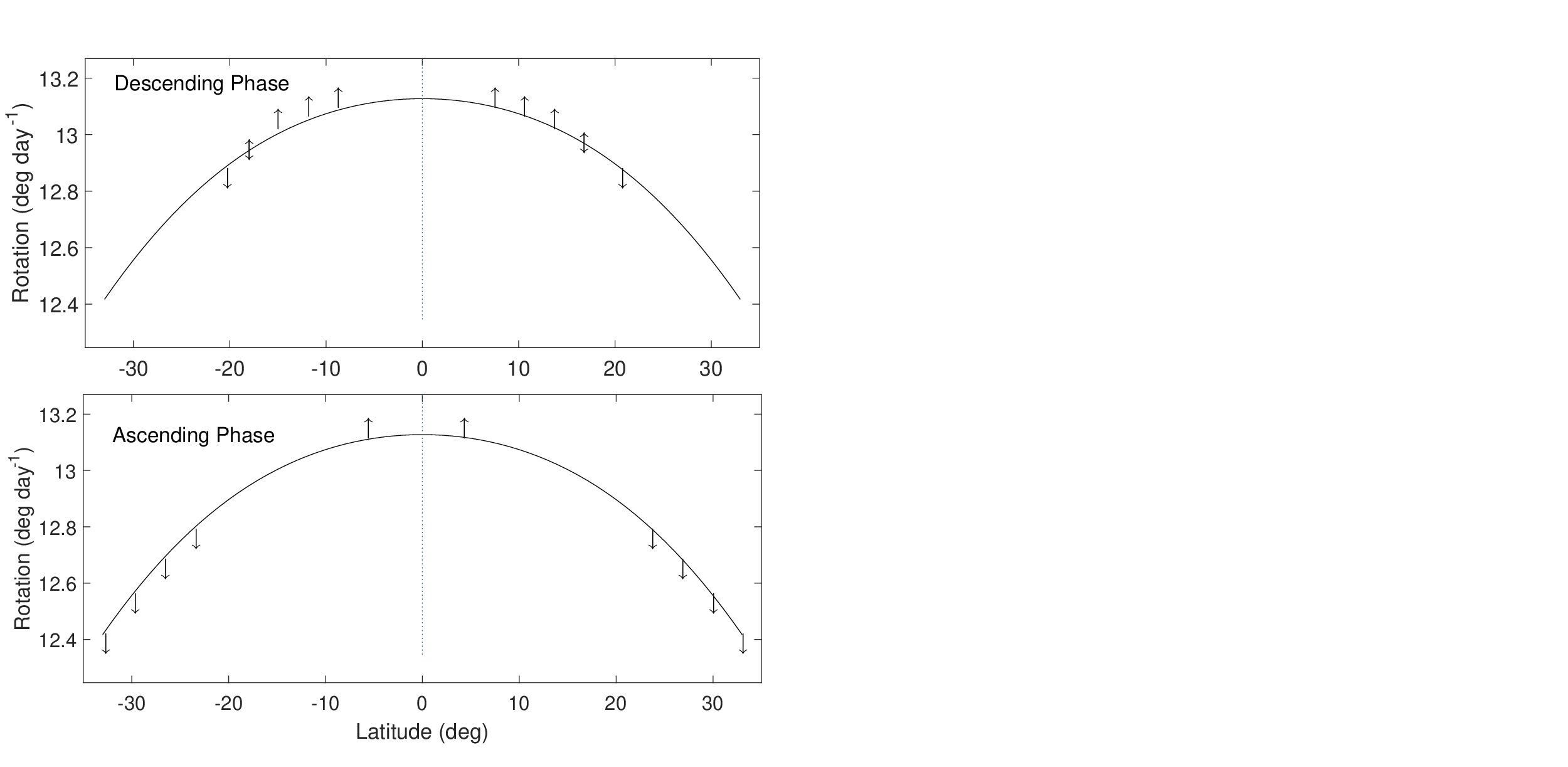}
\caption{Schematic diagram of the effect (up or down arrows) of sunspots on the differential rotation of the quiet chromosphere (the solid line) respectively in the descending (the top panel) and ascending (the bottom panel) phases. Up arrows mean that sunspots increase the rotation velocity of the quiet chromosphere and down arrows mean that sunspots decrease the rotation velocity.
}\label{}
\end{center}
\end{figure}

\section*{Acknowledgments}
We thank the anonymous referee very much  for careful reading of the manuscript  and constructive comments which significantly improved the original version of the manuscript.
This work is supported by Yunnan Fundamental Research Project (202201AS070042, 202101AT070063), the National Natural Science Foundation of China (12373059, 12373061, 11973085, 41964007), the Yunling-Scholar Project (the Yunnan Ten-Thousand Talents Plan), the national project for large scale scientific facilities (2019YFA0405001), the "Yunnan Revitalization Talent Support Program" Innovation Team Project, the project supported by the specialized research fund for state key laboratories, and the Chinese Academy of Sciences.
%the Yunnan Fundamental Research Projects (grant No. 202301AV070007)

\section*{DATA AVAILABILITY}
Synoptic maps of  intensity of the He I 10830\AA\, line from Carrington rotations 2032 to 2135 come from the routine observations of the full-disk chromosphere in the line at the National Solar Observatory (NSO)/Kitt Peak. They are publicly available from the  NSO's web site, ftp://nispdata.nso.edu/kpvt/synoptic/.
% NSO/Kitt Peak data used here are produced cooperatively by NSF/NOAO, NASA/GSFC, and NOAA/SEL.

\clearpage
\end{document}